\documentclass[
aps,pra,reprint,twocolumn,preprintnumbers,superscriptaddress,
amsfonts,amssymb,amsmath,
floatfix,longbibliography]{revtex4-1}

\usepackage{graphicx} 
\usepackage[separate-uncertainty=true, multi-part-units=single]{siunitx} 
\usepackage{color} 
\usepackage{mathtools} 

\usepackage[english]{babel}
\usepackage[
    colorlinks=true,
    frenchlinks=false,
    linkcolor=blue,
    anchorcolor=blue,
    citecolor=blue,
    filecolor=blue,
    urlcolor=blue,
    bookmarks=true,
    bookmarksopen=true,
    bookmarksnumbered=true,
    bookmarksopenlevel=1,
    plainpages=false,
    pdfpagelabels=true,
    draft=false]{hyperref}
\DeclareSIUnit\dBm{dBm}
\DeclareSIUnit\sccm{sccm}
\DeclareSIUnit\minute{min}
\DeclareSIUnit\literSmall{l}
\DeclareSIUnit\bar{bar}

\begin{document}

\title{Cryogenic microwave link for quantum local area networks}

\author{W.~K.~Yam}
\email{WunKwan.Yam@wmi.badw.de}
\affiliation{Walther-Mei{\ss}ner-Institut, Bayerische Akademie der Wissenschaften, 85748 Garching, Germany}
\affiliation{School of Natural Sciences, Technical University of Munich, 85748 Garching, Germany}

\author{M.~Renger}
\altaffiliation{W. K. Y., M. R., and S. G. contributed equally}
\affiliation{Walther-Mei{\ss}ner-Institut, Bayerische Akademie der Wissenschaften, 85748 Garching, Germany}
\affiliation{School of Natural Sciences, Technical University of Munich, 85748 Garching, Germany}

\author{S.~Gandorfer}
\altaffiliation{W. K. Y., M. R., and S. G. contributed equally}
\affiliation{Walther-Mei{\ss}ner-Institut, Bayerische Akademie der Wissenschaften, 85748 Garching, Germany}
\affiliation{School of Natural Sciences, Technical University of Munich, 85748 Garching, Germany}

\author{F.~Fesquet}
\affiliation{Walther-Mei{\ss}ner-Institut, Bayerische Akademie der Wissenschaften, 85748 Garching, Germany}
\affiliation{School of Natural Sciences, Technical University of Munich, 85748 Garching, Germany}

\author{M.~Handschuh}
\affiliation{Walther-Mei{\ss}ner-Institut, Bayerische Akademie der Wissenschaften, 85748 Garching, Germany}
\affiliation{School of Natural Sciences, Technical University of Munich, 85748 Garching, Germany}

\author{K.~E.~Honasoge}
\affiliation{Walther-Mei{\ss}ner-Institut, Bayerische Akademie der Wissenschaften, 85748 Garching, Germany}
\affiliation{School of Natural Sciences, Technical University of Munich, 85748 Garching, Germany}

\author{F.~Kronowetter}
\affiliation{Walther-Mei{\ss}ner-Institut, Bayerische Akademie der Wissenschaften, 85748 Garching, Germany}
\affiliation{School of Natural Sciences, Technical University of Munich, 85748 Garching, Germany}
\affiliation{Rohde \& Schwarz GmbH \& Co. KG, 81671 Munich, Germany}

\author{Y.~Nojiri}
\affiliation{Walther-Mei{\ss}ner-Institut, Bayerische Akademie der Wissenschaften, 85748 Garching, Germany}
\affiliation{School of Natural Sciences, Technical University of Munich, 85748 Garching, Germany}

\author{M.~Partanen}
\affiliation{Walther-Mei{\ss}ner-Institut, Bayerische Akademie der Wissenschaften, 85748 Garching, Germany}

\author{M.~Pfeiffer}
\affiliation{Walther-Mei{\ss}ner-Institut, Bayerische Akademie der Wissenschaften, 85748 Garching, Germany}
\affiliation{School of Natural Sciences, Technical University of Munich, 85748 Garching, Germany}

\author{H.~van der Vliet}
\affiliation{Oxford Instruments NanoScience, Tubney Woods, Abingdon, Oxon, OX13 5QX, UK}

\author{A.~J.~Matthews}
\affiliation{Oxford Instruments NanoScience, Tubney Woods, Abingdon, Oxon, OX13 5QX, UK}

\author{J.~Govenius}
\affiliation{VTT Technical Research Centre of Finland Ltd. \& QTF Centre of Excellence, P.O. Box 1000, 02044 VTT, Finland.}

\author{R.~N.~Jabdaraghi}
\affiliation{VTT Technical Research Centre of Finland Ltd. \& QTF Centre of Excellence, P.O. Box 1000, 02044 VTT, Finland.}

\author{M.~Prunnila}
\affiliation{VTT Technical Research Centre of Finland Ltd. \& QTF Centre of Excellence, P.O. Box 1000, 02044 VTT, Finland.}

\author{A.~Marx}
\affiliation{Walther-Mei{\ss}ner-Institut, Bayerische Akademie der Wissenschaften, 85748 Garching, Germany}

\author{F.~Deppe}
\affiliation{Walther-Mei{\ss}ner-Institut, Bayerische Akademie der Wissenschaften, 85748 Garching, Germany}
\affiliation{School of Natural Sciences, Technical University of Munich, 85748 Garching, Germany}
\affiliation{Munich Center for Quantum Science and Technology (MCQST), 80799 Munich, Germany}

\author{R.~Gross}
\email{Rudolf.Gross@wmi.badw.de}
\affiliation{Walther-Mei{\ss}ner-Institut, Bayerische Akademie der Wissenschaften, 85748 Garching, Germany}
\affiliation{School of Natural Sciences, Technical University of Munich, 85748 Garching, Germany}
\affiliation{Munich Center for Quantum Science and Technology (MCQST), 80799 Munich, Germany}

\author{K.~G.~Fedorov}
\email{Kirill.Fedorov@wmi.badw.de}
\affiliation{Walther-Mei{\ss}ner-Institut, Bayerische Akademie der Wissenschaften, 85748 Garching, Germany}
\affiliation{School of Natural Sciences, Technical University of Munich, 85748 Garching, Germany}
\affiliation{Munich Center for Quantum Science and Technology (MCQST), 80799 Munich, Germany}

\begin{abstract}
Scalable quantum information processing with superconducting circuits is expected to advance from individual processors located in single dilution refrigerators to more powerful distributed quantum computing systems. The realization of hardware platforms for quantum local area networks (QLANs) compatible with superconducting technology is of high importance in order to achieve a practical quantum advantage. Here, we present a fundamental prototype platform for a microwave QLAN based on a cryogenic link connecting two separate dilution cryostats over a distance of $\SI{6.6}{\meter}$ with a base temperature of $\SI{52}{\milli \kelvin}$ in the center. Superconducting microwave coaxial cables are employed to form a quantum communication channel between the distributed network nodes. We demonstrate the continuous-variable entanglement distribution between the remote dilution refrigerators in the form of two-mode squeezed microwave states, reaching squeezing of $\SI{2.10 \pm 0.02}{\decibel}$ and negativity of $\SI{0.501 \pm 0.011}{}$. Furthermore, we show that quantum entanglement is preserved at channel center temperatures up to $\SI{1}{\kelvin}$, paving the way towards microwave quantum communication at elevated temperatures. Consequently, such a QLAN system can form the backbone for future distributed quantum computing with superconducting circuits.
\end{abstract}



\maketitle

\section{Introduction}

Quantum networks are used to provide coherent exchange of quantum states between remote parties~\cite{Kimble2008}. At optical carrier frequencies, several paradigmatic applications of quantum networks have been realized, such as multimode entanglement swapping~\cite{Pompili2021}, long-distance entanglement distribution~\cite{vanLeent2022}, and satellite-to-ground quantum key distribution~\cite{Liao2017}. Meanwhile, superconducting quantum processors operating at microwave frequencies are among the most promising hardware platforms for implementing quantum computing applications due to major breakthroughs demonstrating quantum advantage~\cite{Arute2019} and quantum error correction beyond the break-even point~\cite{Sivak2023, Ni2023}. However, in order to achieve a quantum advantage in practical problems, the number of physical superconducting qubits needs to reach multiple millions~\cite{Webber2022}. This goal will be extremely hard to achieve within the limited experimental space provided by modern cryostats because of their spatial and cooling constraints, as well as being subject to other drawbacks caused by the growing qubit crosstalk on large quantum chips. Whilst massive refrigeration platforms have been proposed~\cite{Hollister2024} as a potential remedy, there may be benefits in a modular approach for large-scale applications towards quantum information processing with superconducting circuits, such as the ability to incrementally expand a network without needing to change or upgrade a central refrigeration plant. Such a solution would employ a large number of smaller individual quantum computing units coherently linked together within a quantum local area network (QLAN)~\cite{Penas2022, Setiawan2023, Rosenblum2018, Zhong2021, VanMeter2016, Awschalom2021}. Low efficiencies of conversion between microwave and optical carrier frequencies~\cite{Sun2006, Mirhosseini2020, Forsch2020} do not allow us to rely on conventional optical methods for communication between remote superconducting qubits. An alternative solution is to employ direct quantum communication at microwave frequencies. Here, recent experiments with propagating quantum microwaves have demonstrated the feasibility of quantum communication in the gigahertz range~\cite{Zhong2021, Fedorov2021, Pogorzalek2019, Magnard2020, Storz2023}. Related experimental protocols employ various superconducting transmission lines as quantum communication channels. These transmission lines are typically cooled to millikelvin temperatures in order to minimize the influence of thermal noise on the quantum properties of propagating microwave signals. By connecting remote dilution cryostats with a cryogenic link containing such superconducting transmission lines, one can realize a quantum link for fundamental Bell tests~\cite{Storz2023} or towards building a prototype microwave QLAN.

\begin{figure*}[!t]
	\begin{center}
		\includegraphics[width=0.9\linewidth,angle=0,clip]{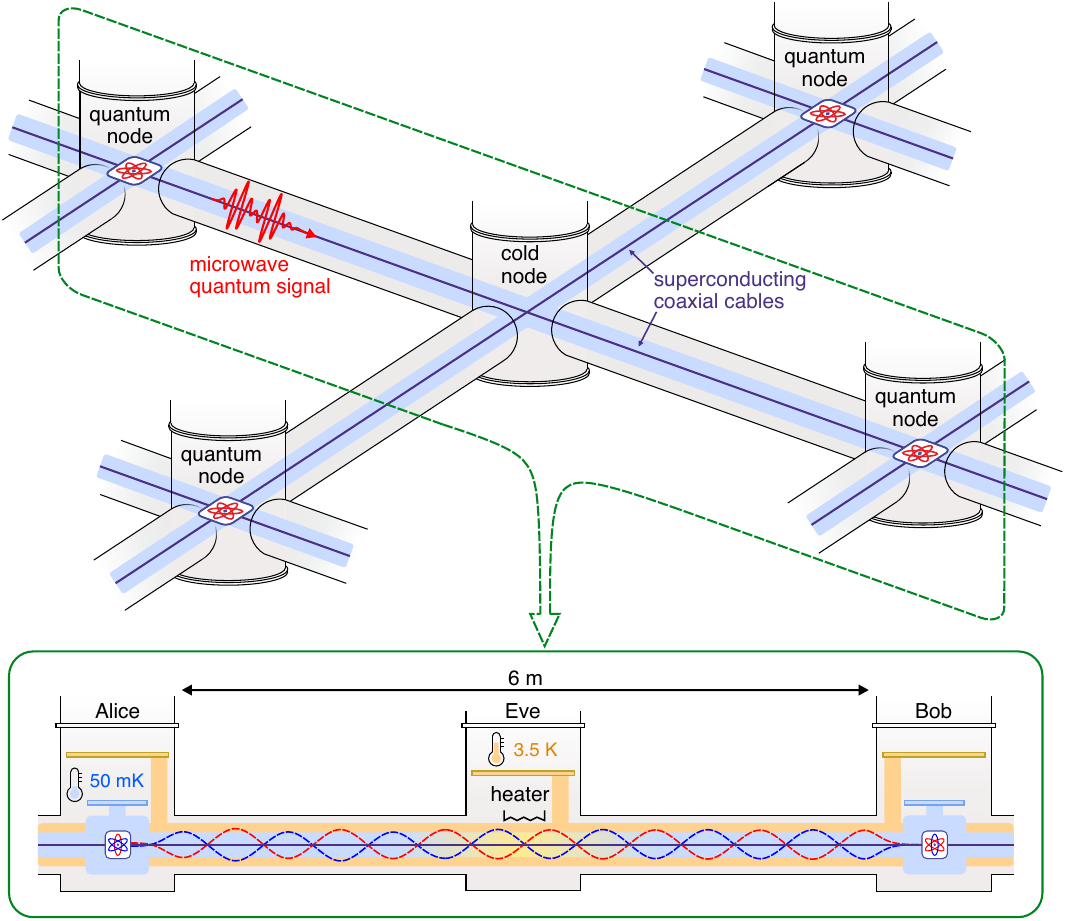}
	\end{center}
	\caption{Square lattice microwave QLAN. Four superconducting quantum nodes within one plaquette are connected to each other via superconducting coaxial cables through respective cryogenic link arms. The bottom panel provides a more detailed schematic of the developed cryogenic link consisting of two quantum nodes (``Alice'' and ``Bob'') and an intermediate cold node (``Eve''), which are used for demonstration of a squeezed state transfer and entanglement distribution at microwave frequencies. Eve does not have a dilution unit and is used to assist cooling for the outer two layers of the cryolink. Local heating can be applied to the center part of the superconducting cables to test fundamental limits of quantum correlation propagation through a thermal environment.}
	\label{Fig:Fig1}
\end{figure*}

In this paper, we present the realization of a $6.6$-meter-long cryogenic link (``cryolink'') connecting two dry dilution refrigerators, which act as quantum communication nodes (referred to as ``Alice'' and ``Bob''). The center of the link consists of a cold node (referred to as ``Eve''). Eve's design allows for connection of additional link arms and a further extension of our microwave QLAN prototype to a square lattice configuration (see Fig.\,\ref{Fig:Fig1}). We implement a $6$-meter-long microwave quantum channel in the innermost stage of the cryogenic link by using three superconducting NbTi coaxial cables. Due to negligibly small thermal conductivity of the superconducting material of these cables and their (deliberately) weak thermal anchoring, we can apply a strong local heating on the cables at the cryolink center without a drastic impact on the base temperatures of Alice and Bob. Hence, the superconducting transmission line through our cryogenic link represents a microwave quantum channel with a variable thermal background. We study this microwave communication channel for transfer of propagating squeezed states and distribution of continuous-variable quantum entanglement. Furthermore, we investigate the impact of the cryolink center temperature $T_\textrm{center}$ on the quantum properties of propagating microwave states. We experimentally distribute path-entangled squeezed microwave states between Alice and Bob with squeezing levels up to $S = \SI{2.10 \pm 0.02}{\decibel}$ and measure negativities up to $N = \SI{0.501 \pm 0.011}{}$. We demonstrate that microwave entanglement distribution is robust against elevated center temperatures of the superconducting cables, up to $T_\textrm{center} = \SI{1}{K}$. This finding indicates that the electromagnetic modes inside the superconducting cables are strongly decoupled from the local thermal baths, in accordance with the fluctuation-dissipation theorem~\cite{Callen1951, Kubo1957}. The propagating microwave states remain robust as long as Alice and Bob temperatures are not affected by heating at the cryolink center and $T_\textrm{center}$ stays sufficiently below the critical temperature of NbTi, $T_\textrm{c} \simeq \SI{9.8}{\kelvin}$, which forms the superconducting cables. As an important technological consequence, preserving quantum coherence in a microwave quantum link does not require millikelvin temperatures along the transmission line, provided a sufficiently high critical temperature and low losses of the superconducting transmission lines. This work demonstrates that microwave cryogenic links can be straightforwardly employed for quantum communication between remote dilution cryostats.

\section{Results}

\subsection{Microwave cryogenic link}

\begin{figure*}[!t]
	\begin{center}
		\includegraphics[width=\linewidth,angle=0,clip]{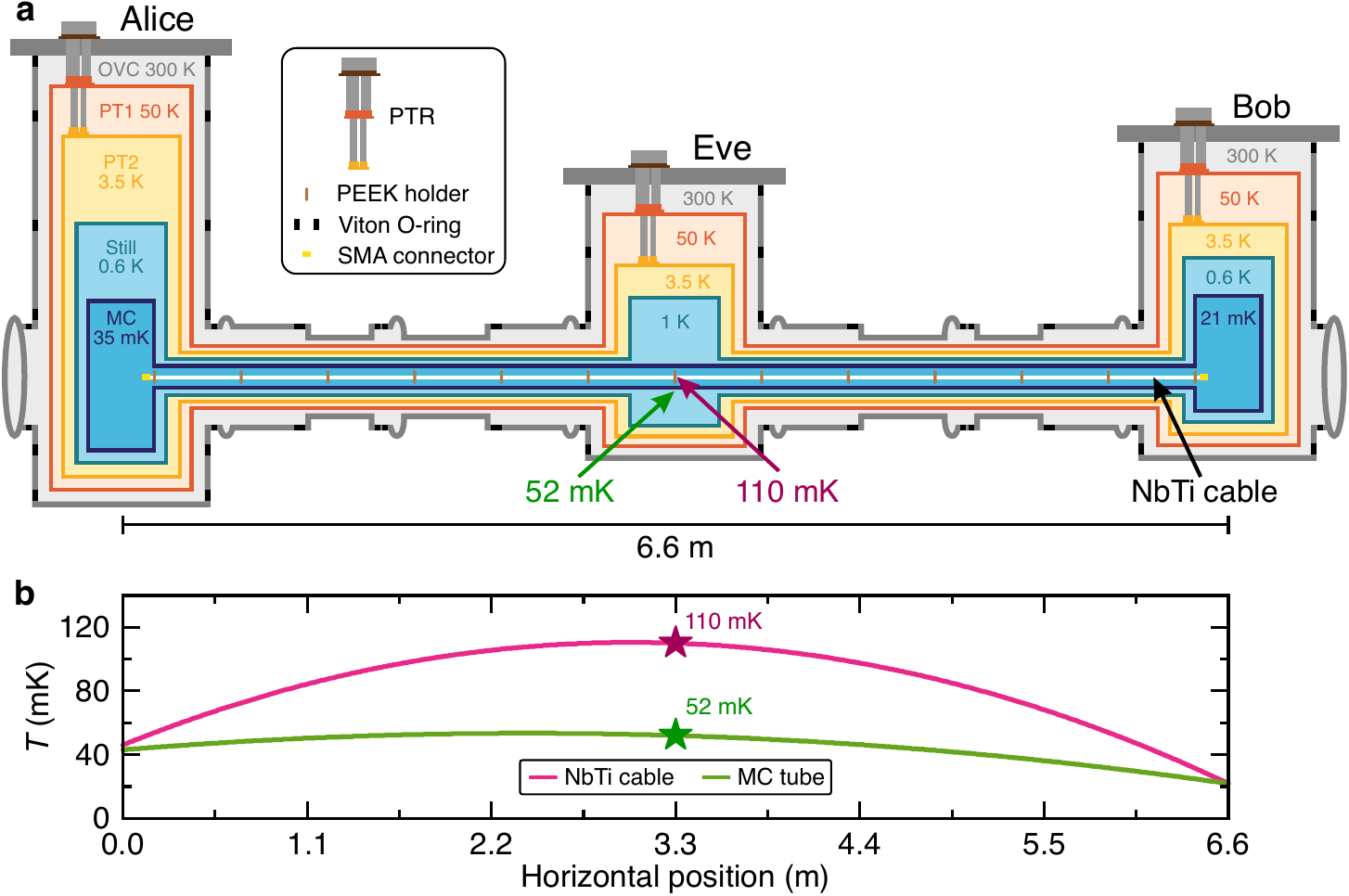}
	\end{center}
	\caption{Layered structure of the cryogenic link. \textbf{a} Cryogenic link schematics consisting of two dry dilution refrigerators (Alice and Bob) spatially separated by $\SI{6.6}{\meter}$. Each stage of the cryogenic link is labeled with its typical steady-state temperature. The innermost stage of the cryogenic link contains three superconducting NbTi coaxial cables, which allow for distribution of microwave quantum signals. A third, intermediate, cryostat without a dilution unit (Eve) is used to support the cooling of the two outermost radiation shields (at temperatures of $\SI{50}{\kelvin}$ and $\SI{3.5}{\kelvin}$, respectively) via an additional pulse tube cooler. \textbf{b} Base temperature profile along the superconducting NbTi cables and MC tube. Stars correspond to measured temperatures at the center of the cryogenic link, for which positions are indicated in panel (\textbf{a}). Solid lines are the result of a heat equation simulation, which takes into account radiation incident on the cables and MC tube, and conductive heat transfer between the cables and MC tube. As boundary conditions, we use the measured temperatures at the MC stages of Alice and Bob.}
	\label{Fig:Fig2}
\end{figure*}

Our microwave cryogenic link consists of two quantum nodes, Alice and Bob, connected by a cold node, Eve. The cryogenic link is designed and assembled in collaboration with Oxford Instruments NanoScience (OINS). Alice is a home-built dry dilution cryostat~\cite{Marx2014} and Bob is a commercial OINS dilution cryostat, both customized to allow a connection with the cryogenic link arm. Eve is also a commercial OINS cryostat, but without a $^3\mathrm{He}/^4\mathrm{He}$ dilution cooling circuit, and is connected to Alice and Bob via two 1.5-meter-long link arm segments. The cryogenic link is vacuum-insulated and includes a layered structure of multiple low-emissivity radiation shields to reduce the absorption of room-temperature radiation. The layered structure of the cryolink coincides with that of the individual Alice and Bob cryostats as schematically illustrated in Fig.\,\ref{Fig:Fig2}a. The radiation shields in the cryolink arms are attached to the respective cryostat shields via semi-cylindrical adapter shells. The link is designed with sufficient mechanical freedom along its symmetry axis to flexibly mount the link arms while ensuring good thermal contact. We use Viton O-ring vacuum seals between room-temperature pieces of the cryolink to attain the background pressure of around $\SI{1e-6}{\milli\bar}$ inside the cryolink. Inside the link arms, radiation shields corresponding to different temperature stages are thermally decoupled from each other via fiberglass spacers. The entire system is designed with multiple degrees of freedom to enable \textit{in-situ} adjustment of relative orientations between Alice, Eve, Bob, and the link arms. Photographs and more details regarding the cryolink system design are provided in Supplementary Note\,1.

A complete cooldown of the cryogenic link takes approximately $\SI{80}{}$ hours. As shown in Fig.\,\ref{Fig:Fig2}a, a pulse tube refrigerator (PTR) is used in each of the cryostats to reach temperatures around \SI{50}{\kelvin} and \SI{3.5}{\kelvin} at the two outermost radiation shield stages. We pre-cool the inner stages to temperatures around $\SI{10}{\kelvin}$ by employing separate helium circuits in Alice and Bob, as well as a helium-based heat switch in Eve~\cite{Poole2023}. Following the pre-cooling phase, $^3\mathrm{He}/^4\mathrm{He}$ dilution cooling circuits in Alice and Bob are used to reach temperatures below \SI{50}{\milli\kelvin} at the innermost mixing chamber (MC) stages. Condensation of the in-flowing $^3\mathrm{He}/^4\mathrm{He}$ mixture is realized via a dry 1-K pot in Alice~\cite{Marx2014} and a Joule-Thomson cryocooler in Bob~\cite{Batey2009}. After the complete cooldown procedure, we reach base temperatures of $\SI{35}{\milli\kelvin}$ at the Alice MC stage, $\SI{21}{\milli\kelvin}$ at the Bob MC stage, and $\SI{52}{\milli\kelvin}$ at the center of the MC tube. Figure\,\ref{Fig:Fig2}b shows the simulated temperature profile across the cryolink. Typical steady-state temperatures of the cryostat stages are indicated in Fig.\,\ref{Fig:Fig2}a. Further details about the cooldown procedure and system performance are provided in Supplementary Note\,2.

The microwave quantum channel is realized using long superconducting NbTi coaxial cables, located within the MC tube of the cryolink, that connect the experimental setups in Alice and Bob. We employ three $6$-meter-long transmission lines, each formed by joining three $2$-meter-long NbTi cables. The MC tube has an inner diameter of $\SI{5.2}{\centi \meter}$ and is attached to the MC shields of Alice and Bob. As illustrated in Fig.\,\ref{Fig:Fig2}a, the NbTi cables are supported by PEEK holders along the cryolink axis and, thereby, thermally decoupled from the MC tube. The superconducting cables are thermally anchored to the Alice and Bob MC stages with silver wires that run along the cables. At the center of the cables, we measure a steady-state temperature of $\SI{110}{\milli \kelvin}$. This elevated temperature, as compared to the MC tube, results from the interplay between the deliberately weak thermal anchoring of the superconducting cables and the external heat-leaks into the system. Potential sources of external heat-leaks are discussed in Supplementary Note 1. An important property of the superconducting cables is their attenuation $\alpha$ per unit length. We measure $\alpha = \SI{1.01}{\decibel \per \kilo\meter}$ at the carrier frequency of $\SI{5.65}{\giga\hertz}$, demonstrating negligibly small dielectric losses. These low losses are necessary to preserve quantum properties of our propagating microwave states. Supplementary Note\,3 provides more technical information about the superconducting transmission lines and loss measurements.

We implement a controlled heating in the cryogenic link by exploiting a weak thermal coupling between the superconducting coaxial cables and the MC stages of Alice and Bob. A proportional–integral–derivative (PID) controllable heater is installed at the center of these cables. Due to the low thermal conductivity of the superconducting NbTi material, the center temperature, $T_\mathrm{center}$, of the cables can be stabilized at elevated local temperatures without a strong effect on the MC temperatures of Alice and Bob. Thus, we can experimentally transfer microwave quantum states across a thermal channel while preserving the operation of superconducting quantum circuits at the endpoints.

\subsection{Squeezed state transfer and entanglement distribution}
\label{Sec:EntanglementDistributionExperiment}

\begin{figure*}[!t]
	\begin{center}
		\includegraphics[width=\linewidth,angle=0,clip]{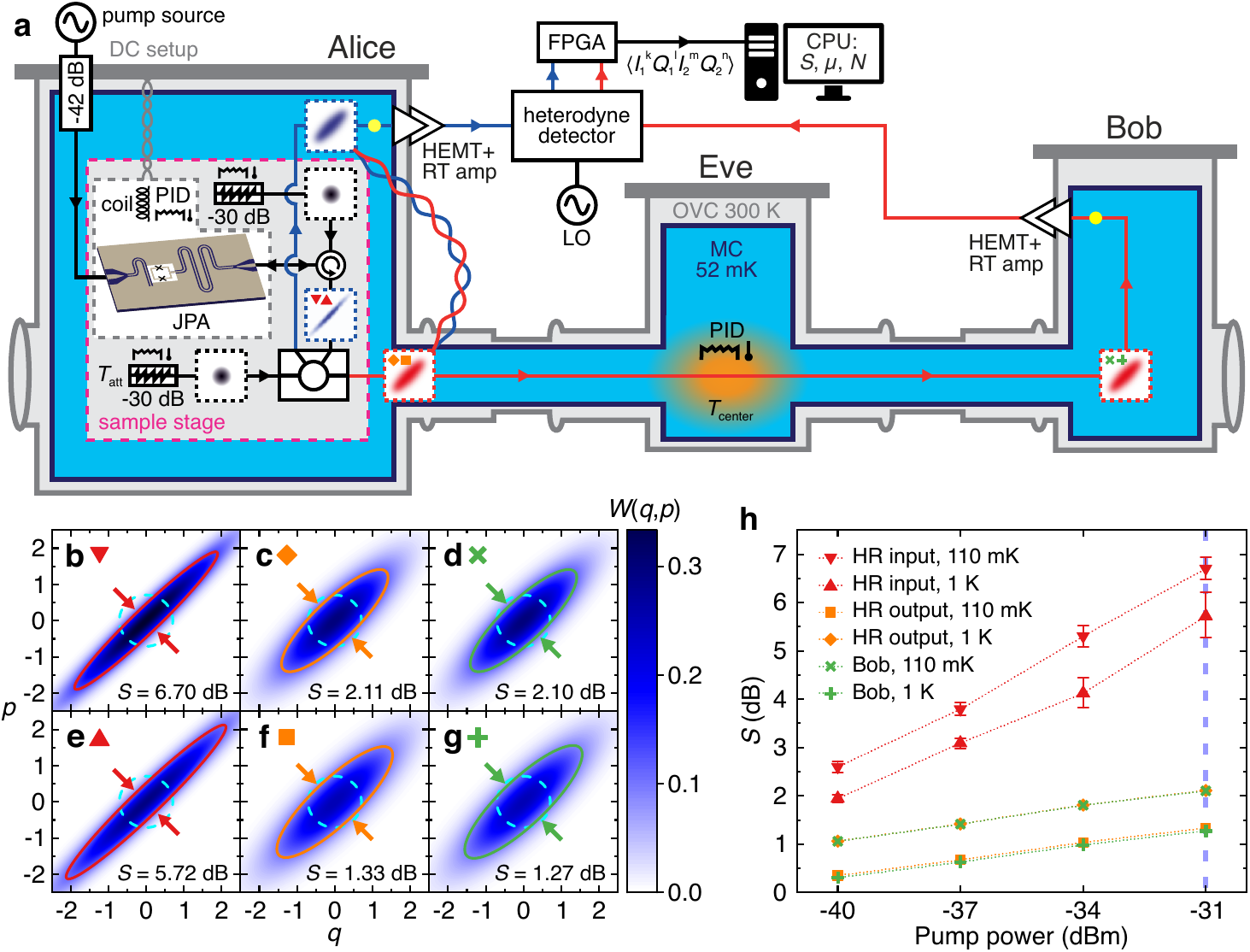}
	\end{center}
	\caption{Microwave squeezed state generation and transfer through the cryolink. \textbf{a} Schematics of the experimental setup in the cryogenic link. Alice employs a flux-driven JPA, which performs squeezing on a weak thermal input state. The resulting squeezed state (red triangle markers) is split at a hybrid ring, which acts as a balanced microwave beam splitter, and the two output modes of the hybrid ring are path-entangled. The output modes (orange square markers) are broadened by weak thermal fluctuations with temperature $T_\mathrm{att}$ incident to the second hybrid ring input port. One of the output modes is kept locally in Alice, while the second mode is transferred over the cryogenic link to Bob (green cross markers). Both modes are reconstructed with joint tomography measurements, and the yellow dots mark the reconstruction points in the Alice and Bob cryostats, respectively. We repeat this experiment for multiple center temperatures of the superconducting cable, from $T_\mathrm{center} = \SI{110}{\milli\kelvin}$ up to $T_\mathrm{center} = \SI{1}{\kelvin}$. \textbf{b} Wigner function of an exemplary squeezed state at the hybrid ring input for $T_\mathrm{center}=\SI{110}{\milli\kelvin}$. Here, the solid line represents the corresponding $1/e$ contour, featuring variance squeezing below the vacuum level indicated by the dashed contour. \textbf{c} Wigner function of the respective squeezed state at the hybrid ring output. \textbf{d} Wigner function of the respective transferred squeezed state at the Bob cryostat. \textbf{e}-\textbf{g} Exemplary Wigner functions for the identical sequence of squeezed states as in panels \mbox{(\textbf{b}-\textbf{d})} for the elevated center temperature, $T_\mathrm{center}=\SI{1}{\kelvin}$. \textbf{h} Squeezing levels $S$ of different states as a function of the pump power at the JPA sample holder. The vertical dashed line denotes the exemplary squeezed states that are displayed in panels \mbox{(\textbf{b}-\textbf{g})}, which are labeled by the respective legend symbols. Dotted lines connecting data points are guides to the eye. When not shown, error bars are smaller than the symbol size.}
	\label{Fig:Fig3}
\end{figure*}

Generally speaking, quantum state transfer (QST) represents a direct transport of quantum information between quantum nodes through a corresponding quantum channel~\cite{Cirac1997}. This goal can be achieved by the exchange of quantum states encoded in propagating single-photon excitations, Gaussian states, magnons, phonons, among others~\cite{Lounis2005, Weedbrook2012, Yuan2022, Bienfait2019}. QST is also related to entanglement distribution, either directly or through more elaborate protocols, such as quantum teleportation or entanglement swapping~\cite{Bennett1993, Żukowski1993}. Here, we focus on QST in the microwave domain, which is relevant for superconducting quantum circuits. It has been experimentally demonstrated for pulsed single-photon, non-Gaussian microwave states~\cite{Kurpiers2018, Bienfait2019}. On the other hand, QST can also work in the steady-state with Gaussian protocols, as demonstrated with squeezed and coherent states~\cite{Pogorzalek2019, Fedorov2021}. These protocols offer substantial advantages over direct QST, as they are less sensitive to communication channel losses and noise. Furthermore, they can potentially be extended to the transfer of arbitrary non-Gaussian states by exploiting pre-distributed two-mode squeezing. This advancement would be a substantial step towards universal QLANs and distributed quantum computing.

In this context, we employ our cryogenic link to realize the direct transfer of squeezed states over a distance of \SI{6.6}{\meter} and the distribution of continuous-variable quantum entanglement in the form of two-mode squeezing. Figure\,\ref{Fig:Fig3}a schematically depicts our experimental setup. At the MC stage of Alice, a flux-driven Josephson parametric amplifier (JPA) is used to generated the squeezed states. This JPA is fabricated at VTT Technical Research Centre of Finland Ltd and consists of a superconducting $\lambda/4$ resonator terminated by a dc-SQUID to the ground. The dc-SQUID provides a Josephson nonlinearity in the resonator and enables flux-tuning of the JPA resonance frequency $\omega_0$~\cite{Wallquist2006}. For our experiment, we choose the microwave signal frequency $\omega_0/2\pi = \SI{5.65}{\giga\hertz}$. By applying a strong coherent pump tone with frequency $\omega_\mathrm{p}$, satisfying the three-wave mixing condition $\omega_\mathrm{p} = 2\omega_0$, the JPA performs a squeezing operation on an input state~\cite{Yamamoto2008, Zhong2013, Yamamoto2016}. We measure the JPA in reflection, where the input and the output states are separated by a cryogenic circulator (see Fig.\,\ref{Fig:Fig3}a). In order to calibrate the photon number of our generated states, we include a $\SI{30}{\decibel}$ attenuator with PID-controllable temperature in the JPA input line for Planck spectroscopy~\cite{Mariantoni2010}. More details about the samples used in this paper are available in Supplementary Note\,4.

In our experiment, the JPA squeezes weak thermal states coming from the \SI{30}{\decibel} attenuator with temperature $T_\mathrm{att}$. The resulting propagating squeezed states are sent to a hybrid ring, which functions as a balanced microwave beam splitter and produces path-entangled, squeezed output states (see Fig.\,\ref{Fig:Fig3}a). One of the output modes is kept in the Alice cryostat while the second output mode propagates through a superconducting cable to the Bob cryostat, thereby accomplishing entanglement distribution. These states are amplified by respective cryogenic high electron mobility transistor (HEMT) amplifiers and subsequently by multiple room-temperature (RT) amplifiers. The amplified signals at the carrier frequency of \SI{5.65}{\giga\hertz} are then down-converted to $\SI{11}{\mega \hertz}$ in a heterodyne detection scheme, similar to the ones used in Refs.~\cite{Pogorzalek2019, Renger2021}. The resulting $\SI{11}{\mega \hertz}$ signals are digitally demodulated and filtered in a field-programmable gate array (FPGA), which enables detection of the statistical quadrature moments $\langle I_1^k Q_1^l I_2^m Q_2^n \rangle$~\cite{Fedorov2021}. For ideal Gaussian states, first-order moments represent the mean and second-order moments represent the variance, while all higher order moments are zero~\cite{Braunstein2005}. We use the statistical moments up to second order to fully reconstruct the Wigner functions in the Gaussian approximation and check for Gaussianity by analyzing the moments up to fourth order. Figures\,\ref{Fig:Fig3}\mbox{b-g} show exemplary Wigner functions of the microwave squeezed states. These Wigner functions correspond to the states at each step of the QST procedure, as indicated by the respective colored markers in Fig.\,\ref{Fig:Fig3}a. We use the reconstructed tomograms to determine quantum state parameters, such as squeezing levels and entanglement monotones~\cite{Perelshtein2022, Menzel2012}.

Squeezing level $S$ is defined as the relative suppression of the squeezed variance $\sigma_\mathrm{s}^2$ below vacuum, $S = -10 \log_{10}\left( \sigma_\mathrm{s}^2/0.25 \right)$. Figure\,\ref{Fig:Fig3}h shows the squeezing levels generated by the JPA at different pump tone powers. We measure maximal squeezing $S = \SI{6.70 \pm 0.23}{\decibel}$ at the hybrid ring (HR) input and successfully transfer squeezed states with $S = \SI{2.10 \pm 0.02}{\decibel}$ over the cryogenic link to Bob. In our experiment, we achieve maximal squeezing at the pump power of \SI{-31}{\dBm} at the JPA sample holder, after which the squeezing level deteriorates due to a gain-dependent JPA noise and higher-order nonlinearities~\cite{Renger2021, Boutin2017}. The exemplary Wigner functions in Figs.\,\ref{Fig:Fig3}\mbox{b-g} correspond to the states generated at this pump power. The path-entangled, two-mode squeezed states are generated at the outputs of the hybrid ring by superimposing the squeezed states from the JPA and weak thermal states from the second \SI{30}{\decibel} attenuator with temperature $T_\mathrm{att}$. These weak thermal states broaden the respective variances of the output modes by at least half the amount of vacuum fluctuations~\cite{Mariantoni2010}, so the squeezing levels at the hybrid ring outputs are fundamentally bound by $S \leq -10\log_{10} 0.5 \simeq \SI{3}{\decibel}$. Finally, we see in Fig\,\ref{Fig:Fig3}h that $S$ decreases as $T_\mathrm{center}$ is raised from \SI{110}{\milli\kelvin} to \SI{1}{\kelvin}. We find that the change in $S$ between the states at the hybrid ring output and at Bob is much smaller than the total change in $S$ due to the increase in $T_\mathrm{center}$. This indicates that most of the degradation in squeezing level is caused by heating of the local thermal bath in the Alice experimental setup.

\subsection{Thermal quantum channel}

\begin{figure*}
	\begin{center}
		\includegraphics[width=0.9\linewidth,angle=0,clip]{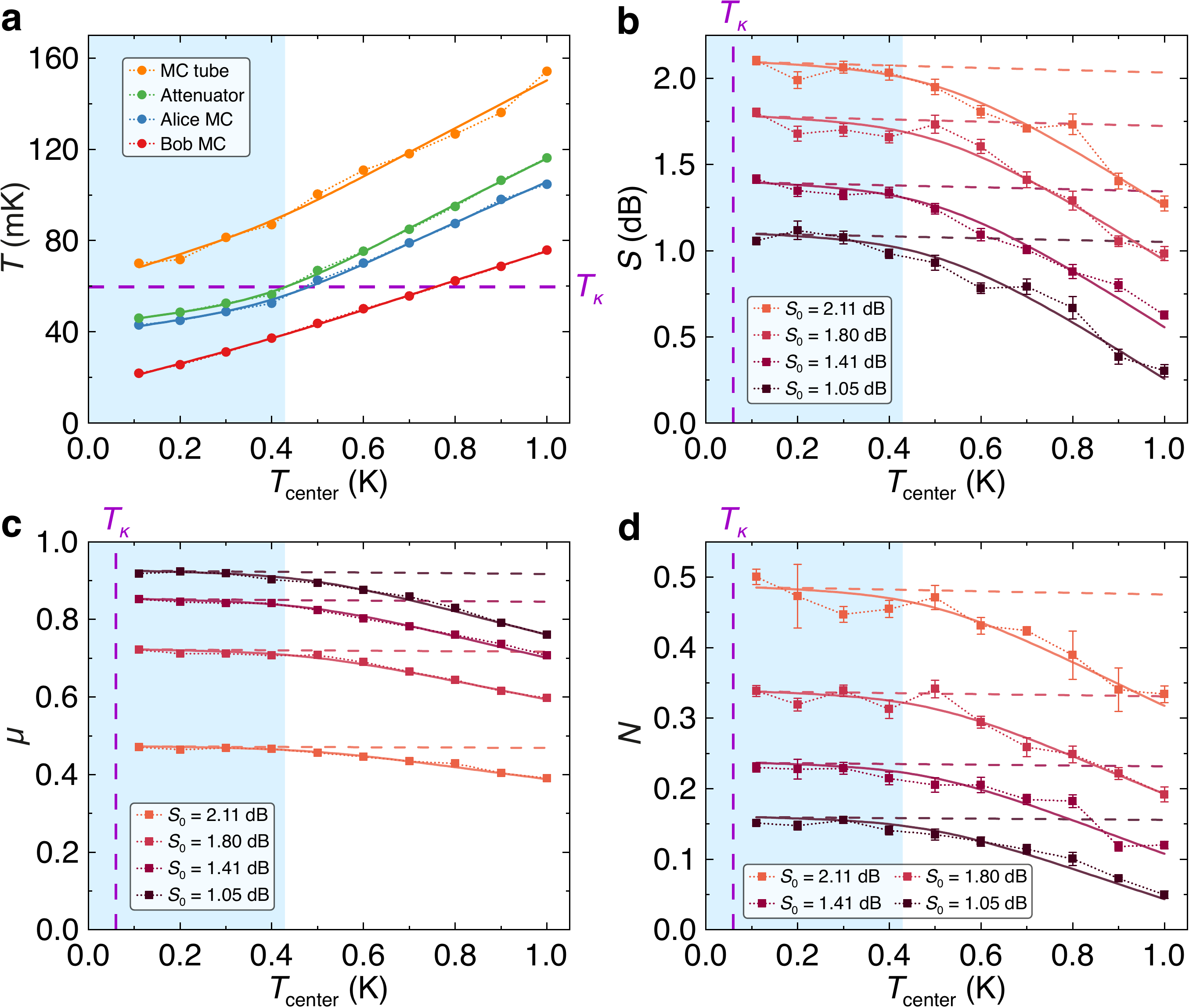}
	\end{center}
	\caption{Squeezed state transfer and entanglement distribution through a thermal microwave channel. \textbf{a} Effect of the superconducting cable center temperature, $T_\mathrm{center}$, on the temperatures $T$ of the MC tube, the \SI{30}{\decibel} attenuators, and the Alice and Bob MC stages. Solid lines represent fits with piecewise linear functions partitioned at $T_\mathrm{center} = \SI{0.4}{\kelvin}$ and smoothed in this region with a sigmoid function. The purple dashed line represents the threshold temperature $T_\kappa = 0.22 \hbar\omega/k_\mathrm{B}$, above which thermal noise becomes noticeable. The light blue color denotes the region where the attenuator temperature is below the threshold temperature, $T_\mathrm{att} \lesssim T_\kappa$, such that noise is considered quantum-limited. \textbf{b} Squeezing level $S$ of transferred states as a function of temperature $T_\mathrm{center}$. Labels $S_0$ correspond to the squeezing levels at the base temperature, $T_\mathrm{center} = \SI{110}{\milli \kelvin}$. The solid lines represent a theory model for the effect of heating in the cryolink center and the subsequent heating in the Alice and Bob MC stages. The dashed lines represent a theory prediction for the case where only the cryolink center is hot, while the Alice and Bob MC stages remain at their base temperatures. \textbf{c} Purity $\mu$ of the transferred squeezed states as a function of $T_\mathrm{center}$. \textbf{d} Entanglement monotone in the form of negativity $N$, between the local Alice mode and the transferred Bob mode, as a function of $T_\mathrm{center}$. We observe $N > 0$ for the entire parameter space, implying the presence of quantum entanglement between the microwave modes in Alice and Bob. Dotted lines connecting data points are guides to the eye. When not shown, error bars are smaller than the symbol size.}
	\label{Fig:Fig4}
\end{figure*}

We use the cryogenic link to implement a tunable thermal channel and repeat the squeezed state transfer and entanglement distribution experiments for various cryolink center temperatures up to $T_\mathrm{center} = \SI{1}{\kelvin}$. This approach enables us to experimentally study the impact of thermal noise on properties of transmitted quantum states. For each $T_\mathrm{center}$ value, we stabilize the cryolink center temperature for approximately one hour until all temperatures in the system reach steady-state before performing measurements. We observe a stray heat transfer over the cryolink when $T_\mathrm{center}$ is raised up to $\SI{1}{\kelvin}$. Figure\,\ref{Fig:Fig4}a shows the steady-state temperatures of various components in the system as a function of $T_\mathrm{center}$. The MC temperatures of Alice and Bob, as well as the MC tube temperature in Eve, show an approximately linear increase as a function of $T_\mathrm{center}$. In Fig.\,\ref{Fig:Fig4}a, we fit the temperature dependence with piecewise linear functions partitioned at $T_\mathrm{center} = \SI{0.4}{\kelvin}$ and smoothed in this region with a sigmoid function. These fits are later used to interpolate temperature values for modeling of the heating impact on propagating quantum microwaves.

Figure\,\ref{Fig:Fig4}b shows the squeezing levels $S$ of the transferred squeezed states as a function of $T_\mathrm{center}$. Each curve corresponds to a fixed JPA pump power and $S_0$ labels the squeezing level at base temperature, $T_\mathrm{center} = \SI{110}{\milli\kelvin}$. We observe that $S$ is relatively insensitive to low $T_\mathrm{center}$, but decreases considerably for higher $T_\mathrm{center}$. We relate this change in temperature dependence to the fact that for sufficiently large $T_\mathrm{center}$, we begin to warm up the Alice MC stage, which leads to an increase in the attenuator temperature $T_\mathrm{att}$. As a result, the JPA input state and the weak thermal state coupled to the second port of the hybrid ring are no longer approximated by vacuum. Instead, these states correspond to thermal quantum mixtures obeying Planck statistics, where each mode on average carries $n_\mathrm{th}(\omega, T) = 1/\{\exp\left[\hbar \omega/(k_\mathrm{B} T)\right] - 1\}$ photons. If we neglect the gain-dependent JPA noise~\cite{Renger2021} and transmission losses, the squeezed variance at the hybrid ring output is given by
\begin{equation}\label{Eq:squeezedVariance}
    \sigma_\mathrm{s}^2 = \frac{1 + 2n_\mathrm{th}(\omega_0, T_\mathrm{att})}{8}\left(e^{-2r} + 1\right),
\end{equation}
where $r$ denotes the squeezing factor of the state generated by the JPA. Detection of output squeezing level $S > 0$ requires $n_\mathrm{th} \leq 1/2$, which implies that $T_\mathrm{att} \leq \hbar \omega_0 /(k_\mathrm{B} \ln3) \approx \SI{0.243}{\kelvin}$.

We model this effect of heating in the cryolink center on the distributed quantum states, as shown in Fig.\,\ref{Fig:Fig4}\mbox{b-d}.  Here, solid lines model the case where propagating microwave signals are affected by the thermal baths from both the heated center of the cryolink and the subsequent heating of the Alice and Bob MC stages. Dashed lines indicate predicted values from our model for the case where only the cryolink center is hot, such that the Alice and Bob MC stages remain at their base temperatures. In our model and its predictions, we consider the input states as weak thermal states defined by their temperatures, which is given by $T_\mathrm{att}$. The propagating quantum states couple to local thermal fluctuations via local losses, as described by the beam splitter model~\cite{Holevo2007, Weedbrook2012}. By comparing our model results with the experimental data in Fig.\,\ref{Fig:Fig4}\mbox{b-d}, we find that the decay in quantum coherence is mainly attributed to the parasitic heating of the Alice and Bob MC stages. The distributed squeezed states appear to be robust against elevated temperatures at the cryolink center due to negligibly small losses of the superconducting coaxial cable. For the model parameters in Fig.\,\ref{Fig:Fig4}\mbox{b-d} (solid and dashed lines), we use relevant MC and attenuator temperatures based on direct measurements (see Fig.\,\ref{Fig:Fig4}a) and estimate losses based on datasheet values. For simplicity, we assume that the temperature across the entire superconducting cable connecting the cryostats is given by its center temperature $T_\mathrm{center}$, which provides us with an upper (pessimistic) bound for the amount of thermal noise coupled to the quantum channel. A more detailed analysis of the temperature profile across the cryolink is given in Supplementary Note\,5.

Next, we investigate threshold temperatures, where thermal noise influences our quantum communication channel. The Planck formula connects the flat, quantum-limited, regime, $n_\mathrm{th}(\omega, T) \to 0$ in the limit $T \to 0$, with the linear, classical, Johnson-Nyquist regime, $n_\mathrm{th}(\omega, T) = k_\mathrm{B}T/(\hbar \omega)$ for $k_\mathrm{B}T \gg \hbar \omega$~\cite{Mariantoni2010}. As a loss-independent quantifier for the transition between these two curvature-free regimes, we take the temperature $T_\kappa = 0.22 \hbar \omega/k_\mathrm{B}$ at which the Planck curve reaches a maximal curvature, satisfying $\partial^3_{T} n_\mathrm{th}(\omega, T) = 0$. For our experimental signal frequency, $\omega_0 = \SI{5.65}{\giga\hertz}$, we find $T_\kappa \approx \SI{59.7}{\milli \kelvin}$. We quantify the quantum regime by $T \lesssim T_\kappa$, since the threshold temperature $T_\kappa$ is more stringent than alternative quantifiers, such as the crossover temperature $T_\mathrm{cr} = \hbar \omega/(2k_\mathrm{B})$~\cite{Mariantoni2010}. For temperatures $T \gtrsim T_\kappa$, the noise spectral density can no longer be regarded as quantum-limited, but rather contains a noticeable thermal contribution. Figure\,\ref{Fig:Fig4}a shows the measured dependence of the attenuator temperature $T_\mathrm{att}$ on $T_\mathrm{center}$, and we observe that $T_\mathrm{att} \lesssim T_\kappa$ for $T_\mathrm{center} \lesssim \SI{0.429}{\kelvin}$ (denoted by a shaded light blue region). At these temperatures, we can simultaneously satisfy $T_\mathrm{att} \lesssim T_\kappa$ and $T_\mathrm{center} \gg T_\kappa$. Hence, we conclude that the thermal bath at the cryolink center does not strongly interact with the transferred squeezed state, apart from its parasitic heating of the Alice and Bob MC stages. This effect becomes more clear in Fig.\,\ref{Fig:Fig4}c, where the purity $\mu$ is approximately independent of $T_\mathrm{center}$ up to $T_\mathrm{center} \simeq \SI{0.429}{\kelvin}$, and correspondingly $T_\mathrm{att} \simeq T_\kappa$. Purity, $\mu$, measures a deviation of the state uncertainty from the Heisenberg limit, hence providing an estimate of quantum statistical mixing with the environment~\cite{Braunstein2005}. The sustained purity up to $T_\mathrm{center} \simeq \SI{0.429}{\kelvin}$ shows that quantum coherence is preserved despite the thermal nature of the transmission channel. We note that purity decreases with squeezing level, which is due to increased gain-dependent JPA noise at higher pump powers.

The squeezed states transferred to the Bob cryostat are path-entangled with the squeezed states kept in the Alice cryostat. To determine the amount of entanglement contained between the local mode in Alice and the transferred mode in Bob, we estimate negativity $N$ using the two-mode covariance matrix that describes the entire two-mode system~\cite{Menzel2012, Fedorov2018}. Negativity is an entanglement monotone based on the necessary and sufficient Peres-Horodecki PPT criterion for separability~\cite{Peres1996}. Positive values of negativity, $N > 0$, imply quantum entanglement between corresponding modes. When neglecting losses and gain-dependent JPA noise, we have the simple relationship, $N = \max\left[(10^{S/10}-1)/2, 0\right]$. Consequently, any positive squeezing levels give rise to entanglement. Figure\,\ref{Fig:Fig4}d shows the experimentally reconstructed negativity, where we reach maximal negativity $N = \SI{0.501 \pm 0.011}{}$. Furthermore, we have $N > 0$ for the entire parameter space, which demonstrates the successful entanglement distribution.

It is notable that even at $T_\mathrm{center} = \SI{1}{\kelvin}$, which corresponds to an average of $\SI{3.3}{}$ thermal photons per mode, we are not affected by the sudden death of entanglement~\cite{Yu2009, Renger2022}. When environmental noise is coupled into a quantum channel, sudden death of entanglement is expected to occur (independently of $S$) when the injected noise exceeds one photon, $n_\mathrm{th}(\omega_0, T_\mathrm{center}) = 1$, corresponding to $T_\mathrm{center} = \hbar \omega_0/(k_\mathrm{B} \ln 2) \approx \SI{0.385}{\kelvin}$. Thus, our observation of entanglement distribution up to $T_\mathrm{center} = \SI{1}{\kelvin}$ experimentally illustrates the fluctuation-dissipation theorem~\cite{Callen1951}. The theorem states that the variance of the voltage fluctuations $\langle V^2 \rangle$ in the transmission line, within the single-side measurement bandwidth $B$, is given by
\begin{equation}\label{Eq:FluctuationDissipation}
    \langle V^2 \rangle = 2\hbar \int_{\omega_0 - B}^{\omega_0 + B} \coth\left(\frac{\hbar \omega}{2k_\mathrm{B}T_\mathrm{center}}\right)\varepsilon(\omega)d\omega,
\end{equation}
where $\varepsilon(\omega)$ corresponds to the dissipation spectrum, determined by the Fourier-transformed voltage susceptibility of the system~\cite{Kubo1957}. For our superconducting cables, the dissipation spectrum is determined by its ohmic resistance. Therefore, we have $\varepsilon(\omega) \ll 1$ for frequencies well below the superconductor gap frequency, $\omega \ll \omega_\Delta$, where $\omega_\Delta/2\pi \simeq \SI{370}{\giga \hertz}$ for NbTi. This extremely small $\varepsilon(\omega)$ strongly suppresses the contribution from the thermal bath with $T_\mathrm{center}$ in Eq.\,(\ref{Eq:FluctuationDissipation}). Moreover, the characteristic thermal energy scale in our experiment is much smaller than the superconductor gap energy $\Delta(T_\mathrm{center})$, so we can neglect the surface resistance $R_\mathrm{s} \propto \exp\left[-\Delta(T_\mathrm{center})/(k_\mathrm{B}T_\mathrm{center})\right]$ of the superconducting cable~\cite{Turneaure1968, Bardeen1957}. Hence, the phononic environmental temperature and photonic mode temperature of the superconducting cable are effectively decoupled from each other. The thermal fluctuation spectrum in the microwave signal line is instead limited by the input mode temperature $T_\mathrm{att}$. Although the fragile quantum states propagate through a hot thermal bath, interaction between the states and the bath are suppressed as long as the cable temperature remains well below the critical temperature of the superconductor.

\section{Discussion}

In summary, we have realized a hardware platform for microwave quantum local area networks by connecting two dilution refrigerators over a distance of $\SI{6.6}{\meter}$ via a cryogenic link, which reaches a center temperature of  $\SI{52}{\milli \kelvin}$. The innermost stage of our cryolink contains three superconducting NbTi coaxial cables, each of length \SI{6}{m} and forming a microwave quantum channel. We reach a typical temperature of $T_\mathrm{center} = \SI{110}{\milli \kelvin}$ at the center of the cables, which results from an interplay between their deliberately weak thermal anchoring and residual heat loads. We provide technical details about the cryolink design and assembly, as well as its operation and performance. We employ our system to establish a thermal quantum channel with a tunable temperature between two spatially-separated quantum nodes. We demonstrate the successful quantum state transfer of squeezed states with squeezing levels up to $S = \SI{2.10 \pm 0.02}{\decibel}$. Additionally, we demonstrate the distribution of continuous-variable microwave entanglement between the remote dilution fridges with negativities up to $N = \SI{0.501 \pm 0.011}{}$. This entanglement distribution remains robust against thermal noise, even when we increase the center temperature of the superconducting cables up to $T_\mathrm{center} = \SI{1}{\kelvin}$. The corresponding thermal noise photon number significantly exceeds the threshold for the sudden death of entanglement. Thus, our results experimentally illustrate the fluctuation-dissipation theorem by showing that environmental noise hardly affects the quantum coherence in the transmission line, provided that it is in the superconducting state. As an important technological consequence, microwave QST and entanglement distribution do not necessarily require millikelvin links, but can also be realized at significantly higher temperatures. Operation at temperatures of liquid helium, or even liquid nitrogen, are possible as long as the corresponding transmission lines remain in the superconducting state and have low absorption losses at microwave frequencies.

As an outlook, the demonstrated cryogenic link can serve as a versatile platform for many fundamental and applied experiments. One of the most straightforward steps is to implement microwave quantum teleportation between remote locations. The QLAN can also be expanded to include many quantum nodes, using our scalable lattice architecture, for fundamental studies of multipartite entanglement. Furthermore, techniques with squeezed microwaves can be used for remote entanglement of superconducting qubits, paving the way towards distributed quantum computing~\cite{Agusti2022, Didier2018} and hybrid quantum information processing~\cite{Andersen2015, Takeda2013}. More explorative experiments include dark matter axion searches~\cite{Lasenby2020, Braine2020}, where the cryolink would allow the separation of large magnetic fields, as required in axion detection schemes, from fragile quantum-limited detectors based on superconducting circuits.

\section{Methods}

\subsection{Superconducting NbTi coaxial cable losses}

The microwave transmission lines inside the cryogenic link consist of 2-meter-long superconducting NbTi coaxial cable segments connected by superconducting joints. Microwave losses in these cables need to be carefully characterized, since strong thermal states due to the local heating in the cryolink couple to propagating quantum signals via the corresponding channel attenuation. We employ resonator type of measurements to determine attenuation in the superconducting cable~\cite{Kurpiers2017}. We find that at our signal frequency of \SI{5.65}{\giga\hertz} the microwave losses are around \SI{1.01}{\decibel\per\kilo\meter}. Supplementary Note\,3 provides more details about the related measurements.

\subsection{Cryogenic link temperature profile}

The 6-meter-long superconducting cables inside the cryogenic link are connected at each end to the respective experimental setups in the Alice and Bob MC stages. We model a temperature profile $T(x)$ along these cables by treating them as a quasi-1D object and using independent temperature measurements in the Alice MC, the Bob MC, and the cyrolink center. At cryogenic temperatures, thermal coupling between the superconducting cables and the MC stages is dominated by the silver wires with thermal conductivity $\lambda_{\mathrm{Ag}}$ and cumulative cross section $A_\mathrm{Ag}$. We use the steady-state heat equation
\begin{equation}
    -\lambda_{\mathrm{Ag}} A_\mathrm{Ag} \frac{\partial^2 T}{\partial x^2} = \dot{Q}_{\mathrm{c}} + \dot{Q}_{\mathrm{r}} + \dot{Q}_{\mathrm{h}},
\end{equation}
where $\dot{Q}_{\mathrm{c}}$ is the conductive heat transfer between the MC tube and the superconducting cable, $\dot{Q}_{\mathrm{r}}$ is the radiative heat load, and $\dot{Q}_{\mathrm{h}}$ is the contribution from the local heating. We numerically solve this heat equation using a finite-difference method, where we use Alice's attenuator temperature and Bob's MC temperature as Dirichlet boundary conditions, to provide results as indicated in Fig.\,\ref{Fig:Fig2}b. Supplementary Note\,5 provides more details about this calculation.

\subsection{Squeezing level, purity and negativity}

Relevant quantum state parameters can be calculated using the $N$-mode quadrature field operators $\mathbf{\hat{R}} = (\hat{I}_1, \hat{Q}_1, \ldots, \hat{I}_N, \hat{Q}_N)^\mathrm{T}$. We employ the vacuum variance definition of $1/4$. Gaussian states are fully described by statistical quadrature moments up to second order. The first moment is the displacement vector $\mathbf{d} = \langle \mathbf{\hat{R}} \rangle$, which is zero for our transferred (two-mode) squeezed states. The second moment is the covariance matrix $\mathbf{V}$, which is defined by~\cite{Weedbrook2012}
\begin{equation}
    V_{ij} = \frac{1}{2} \langle \{ \Delta\hat{R}_i, \Delta\hat{R}_j \} \rangle,
\end{equation}
where $\Delta\hat{R}_i = \hat{R}_i - \langle\hat{R}_i\rangle$ and $\{,\}$ is the anti-commutator. The variance of each quadrature operator is given by the corresponding diagonal element $V_{ii} = \langle (\Delta\hat{R}_i)^2 \rangle = \langle \hat{R}_i^2 \rangle - \langle \hat{R}_i \rangle^2$. Squeezing level $S$ describes the suppression of a state's squeezed variance $\sigma_\mathrm{s}^2$ below the vacuum variance. We find the squeezing level using
\begin{equation}
    S = -10 \log_{10} \left( \frac{\sigma_\mathrm{s}^2}{0.25} \right),
\end{equation}
where $\sigma_\mathrm{s}^2$ is derived from the quadrature variances $V_{ii}$. Purity $\mu$ describes the quantum statistical mixing of a state with its environment. The purity of a Gaussian state is given by
\begin{equation}
    \mu = \frac{1}{\sqrt{16 \det\mathbf{V}}},
\end{equation}
where $\mathbf{V}$ is the covariance matrix associated with the state. Entanglement monotones quantify the amount of entanglement contained in quantum states, including the two-mode squeezed state. One such entanglement monotone is negativity, which is based on the PPT criterion for separability. We can analyze two-mode Gaussian states using the covariance matrix in its block form
\begin{equation}
    \mathbf{V} =
    \begin{pmatrix}
        \mathbf{A} & \mathbf{C} \\
        \mathbf{C}^\mathrm{T} & \mathbf{B}
    \end{pmatrix},
\end{equation}
where $\mathbf{A}$, $\mathbf{B}$, and $\mathbf{C}$ are $2 \times 2$ real matrices, and its symplectic eigenvalues~\cite{Serafini2004}
\begin{equation}
    \nu_\pm = \sqrt{ \frac{\sqrt{\Delta \pm \sqrt{\Delta^2 - 4 \det\mathbf{V}}}}{2} },
\end{equation}
where $\Delta = \det\mathbf{A} + \det\mathbf{B} + 2\det\mathbf{C}$. In particular, the negativity is given by~\cite{Adesso2005}
\begin{equation}
    N = \max\left[ 0, \frac{1 - 4\nu_-}{8\nu_-} \right].
\end{equation}

\subsection{Quantum state transfer and entanglement distribution model}

We model a squeezed state transfer and two-mode entanglement distribution with a sequence of operators acting on the two-mode covariance matrices that describe the corresponding quantum states. We do not consider transformations of the displacement vector, because we only deal with zero-mean Gaussian states in our experiments. The input covariance matrix is
\begin{equation}
    \mathbf{V}_\mathrm{in} =
    \begin{pmatrix}
        (1 + 2n_\mathrm{th}) \mathbf{\mathbf{I}_2} & \mathbf{0}_2 \\
        \mathbf{0}_2 & (1 + 2n_\mathrm{th}) \mathbf{I}_2
    \end{pmatrix},
\end{equation}
where $n_\mathrm{th}$ is the photon number of the input weak thermal states and $\mathbf{I}_2$ ($\mathbf{0}_2$) is the $2 \times 2$ identity (zero) matrix. The squeezing operation of the JPA is described by a noisy squeeze operator $\hat{S}$. The hybrid ring is described by a balanced beam splitter operator $\hat{B}$. The transmission over our $\SI{6.6}{\meter}$ cryogenic link is described by an asymmetric beam splitter operator $\hat{C}$ with transmissivity, $\tau$, linked to the superconducting cable losses in linear units, $\varepsilon = 1 - \tau$. We model microwave losses along specific propagation segments with loss operators $\hat{L}_i$. The output covariance matrix is given by
\begin{gather}
    \mathbf{V}_\mathrm{out} = \hat{U} \mathbf{V}_\mathrm{in} \hat{U}^\dagger, \\
    \hat{U} = \hat{L}_4 \hat{C} \hat{L}_3 \hat{B} \hat{L}_2 \hat{S} \hat{L}_1.
\end{gather}
The propagating quantum states are coupled to local thermal baths via the local loss operators $\hat{L}_i$, as described by the beam splitter model~\cite{Holevo2007, Weedbrook2012}. We use experimentally measured values for the relevant temperatures and corresponding datasheet values of losses to obtain exact numerical values for the loss operators. We fit the measured temperature points in Fig.\,\ref{Fig:Fig4}a using piecewise linear functions partitioned at $T_\mathrm{center} = \SI{0.4}{\kelvin}$ and smoothed in this region with a sigmoid function,
\begin{equation}\label{Eq:temperatureFitting}
    f(T) = a T + b + \frac{1 + \tanh[4(T-0.4)]}{2} (c - a) (T-0.4)
\end{equation}
where $T$ is the varying center temperature in kelvin and $a$, $b$, and $c$ are fitting parameters.

\medskip\noindent
\textbf{\large Data availability}

\noindent
The data that support the findings of this study are available from the corresponding authors upon reasonable request.

\medskip\noindent
\textbf{\large Acknowledgements}

\noindent
We acknowledge valuable support of Kurt Uhlig regarding modifications of our home-built dilution refrigerator. We acknowledge the work of Janne Lehtinen consisting of the supervision of process development in the clean room and room-temperature characterization of the JPAs. We acknowledge Craig Bone for his instrumental work in the build and test of this project, Andy Czajkowski, Anthony Swan, and Tom Marsh for their support during test, Chris Wilkinson for the design of the cryogenic link, Will Dunsby for the welding of the copper MC radiation shield. We acknowledge support by the German Research Foundation via Germany`s Excellence Strategy (EXC-2111-390814868), the EU Quantum Flagship project QMiCS (Grant No.~820505), the German Federal Ministry of Education and Research via the project QUARATE (Grant No.~13N15380). We acknowledge funding from European Union’s Horizon 2020 research and innovation programme under grant agreement No. 824109 European Microkelvin Platform (EMP), the Academy of Finland through the QTF Centre of Excellence project No. 336817, Business Finland through QuTI-project  (No. 128291). We also acknowledge Technology Industries of Finland Centennial Foundation for funding. This research is part of the Munich Quantum Valley, which is supported by the Bavarian state government with funds from the Hightech Agenda Bayern Plus.

\medskip\noindent
\textbf{\large Author contributions}

\noindent
K.G.F. and F.D. planned the experiment. M.R., S.G., and W.K.Y. performed the measurements and analyzed the data. W.K.Y., M.R., S.G., F.F., M.H., K.E.H., F.K., Y.N., M.Pa, M.Pf, A.M., F.D., and K.G.F contributed to the construction of the cryolink. M.Pa. and M.Pf. calibrated the superconducting cable losses. H.vdV. and A.J.M. contributed to the design and assembly of the cryolink. J.G., R.N.J., and M.Pr. provided the JPA samples. K.G.F., F.D., A.M., and R.G. supervised the experimental part of this work. W.K.Y., M.R., S.G., and K.G.F. wrote the manuscript. All authors contributed to discussions and proofreading of the manuscript.

\medskip\noindent
\textbf{\large Competing interests}

\noindent
The authors declare no competing interests.

\bibliography{Bibliography}
\medskip\noindent

\end{document}